# ISO/EPC Addressing Methods to Support Supply Chain in the Internet of Things

Abolfazl Qiyasi Moghadam
Department of Electrical and Computer Engineering,
Faculty of Shahid Shamsipour, Tehran Branch, Technical
and Vocational University (TVU), Tehran, Iran.
a.qiyasimoghadam@gmail.com

Mehdi Imani
Department of Electrical and Computer Engineering,
Faculty of Shahid Beheshti, Alborz Branch, Technical and
Vocational University (TVU), Karaj, Iran.
m.imani@gmail.com

*Abstract*—**RFID systems are among the major infrastructures of the Internet of Things, which follow ISO and EPC standards. In addition, ISO standard constitutes the main layers of supply chain, and many RFID systems benefit from ISO standard for different purposes. In this paper, we tried to introduce addressing systems based on ISO standards, through which the range of things connected to the Internet of Things will grow. Our proposed methods are addressing methods which can be applied to both ISO and EPC standards. The proposed methods are simple, hierarchical, and low cost implementation. In addition, the presented methods enhance interoperability among RFIDs, and also enjoys a high scalability, since it well covers all of EPC schemes and ISO supply chain standards. Further, by benefiting from a new algorithm for long EPCs known as selection algorithm, they can significantly facilitate and accelerate the operation of address mapping.**

## I. Introduction

The future Internet is the Internet of Things, and one of the most challenges of Internet of Things is the way objects interact with each other. One of the infrastructural technologies of the Internet of Things is RFID, which is used for identifying different objects and can remotely save the information in objects through a tag or recover them. We can add the tags to anything: our household objects, products, animals, and even humans. This helps us to have a better control over our environment, so that in case they are lost or stolen, we would be better able to track them. The main challenge faced by RFID today is that tracking and monitoring objects and products should be possible through RFIDs. Further, since RFID systems are among the main infrastructures of the Internet of Things, then the basis of connecting these systems to the Internet should be provided. In order to connect RFID systems to the Internet, first these systems should be given an address, and these addressing methods should comply with the RFID standards. RFID system has different standards including ISO and EPC, which are constantly developing. Various addressing systems have been given for EPC standard, but unfortunately ISO standard has not been considered for over time.

So far, different organizations have been competing for developing RFIDs, and each of them has added some services to RFIDs. Walmart & Department of Defense (DoD) are two important organizations in developing RFIDs, each of which needed standards for developing this system. Auto-ID Center and International Standards Organization (ISO) are two organizations that deal with presenting and developing standards for RFID systems. ISO has provided various standards with different purposes, such as the standards related to air interface, which is the wireless communication circuit between two devices. On the other hand, Auto-ID organization has presented EPC standard which is controlled by EPCglobal. Walmart organization employs EPC standard, but DOD organization benefits from EPC for more general purposes, while it uses ISO for Air Interface [1]. EPC standard has presented different classes (see Table 1), which should match ISO standard. However, it was not this way until eventually EPC standard managed to introduce its second-generation, which matches ISO air interface. The most important ISO standards for RFID systems are ISO 11784, ISO/IEC 18000-1 to 18000-7, ISO/IEC 17363 to 17367 and etc. [3].

All these standards in interaction with each other to support supply chain and provide important services. Supply chain is a multilevel concept, which encompasses all aspects of producing a product from the raw materials to the final product. Five standards have been provided by ISO for supply chain with the help of RFID known as supply-chain applications of RFID (ISO 17363 to 17367). Use of RFID in the supply chain has resulted in development of electronic systems for products. This, in turn, has led to emergence of characteristics such as layering the supply chain, tracking and monitoring the products in each layer or any place, etc. [4].

TABLE I.  EPC CLASSES [2]

| EPC Class Type | Features | Tag Type |
|---|---|---|
| Class 0 | Read Only | Passive (64 bit only) |
| Class 1 | Write Once, Read Many (WORM) | Passive (96 bit min.) |
| Class 2 (Gen2) | Read/ Write | Passive (96 bit min.) |
| Class 3 | Read/Write with battery power to enhance range | Semi-Active |
| Class 4 | Read/Write active transmitter | Active |

One of the most important standards available for RFID system is ISO 15963 presented as Unique Identification of RF Tag [5]. This standard defines a numerical system called RFID tag identifier. The ID that is used for detecting RFID during the communication between RFID and a reader is called Unique ID (UID). The subsequent important standards include ISO 15459 series (ISO 15459 all parts), which work exclusively on Unique Item Identifier (UII) [6 to 11]. ID and UII are located in different sites of the RFID memory, with the structure of this memory being further explained in subsequent sections.

The rest of this paper is organized as follows. Section II deals with investigating related works, which are mostly related to EPC standard. In Section III, the structure of RFID tag memory has been investigated. Section IV studies the proposed method, and eventually the paper is concluded in Section V.

## II. RELATED WORKS

Unfortunately, an addressing method benefiting from ISO standards has not been presented so far, but the methods that work based on EPC are further explained.

Authors in [12] have presented a cryptographic addressing mechanism based on Host Identity Protocol (HIP). This method works with homogeneous tags. HIP obtains the ID of two homogeneous tags and encapsulates it in tables (HIP header) by exchanging messages between them. Host Identifier Tag (HIT) uses a one-way hash function to encrypt the EPC. A Network Address Translator (NAT) converts the HIT values to an IPv6 address and attaches it to each tag.

The protocol proposed in [13] focuses on adaption of IPv6 with sensor networks, which is called Sensor Network for All IP World (SNAIL). This mechanism is compatible with all 6LowPAN features, and it can be implemented in every Personal Area Network (PAN) type such as Inter PAN and Intra PAN. This method covers four features in addition to IP compatibility including: 1- mobility management, which allows for access to the movable devices permanently. SNAIL platform enables the mobility by MARIO protocol (Mobility Management Protocol to Support Intra PAN and Inter PAN with Route Optimization for 6LowPAN) which is based on MIPv6 and reduces the handover delay. 2- web enablement: SNAIL uses HTTP for devices in order to have direct access to the web. 3- Time Synchronization: 6LowPAN Network Time Protocol (6LNTP) is based on Network Time Protocol (NTP), and Simple NTP (SNTP) performs the time synchronization between the nodes in IoT. 4- Security: objects that are connected to the Internet must be secure against various attacks, therefore, SNAIL uses SSL based elliptic curve cryptography (ECC) security algorithms and protocols (e.g. MD5, RC4, ECDH, SHAI 1 and etc.) to secure the end-to-end messages in WSNs.

A hardware based approach is presented in [14]. RFID is equipped with a circuit which handles the mapping process of EPC to IPv6. This system uses IEEE 802.11 (WLAN) to connect to the RFID tags. Also, an EEPROM is attached to the system, which maps the information of addressing. The system eliminates the reader for RFID using Wireless Network Interface (WNIC) to communicate with tags.

Authors in [15] have done an investigation based on Building Automation System and Control Network (BACnet), called BAS (Building Automation System). BACnet applications include Heating, Ventilation, Air Conditioning (HVAC), Lightening, Security, Household Devises and etc. BACnet is suitable for Local Area Networks (LANs) and uses IPv4. BACnet is, however, unable to use IPv6 which confronts it with interoperability and scalability problems due to the device identification. Meanwhile, the integration of IPv6 and BACnet is proposed in BAS and the Human-Things interaction improves by BAS. Therefore, some useful applications of BAS are: 1) device maintenance (e.g. home appliances), with the device itself detecting the system problems reporting them to the operator. 2) energy harvesting and intelligent systems 3) use of BAS in commercial fields like conferences by using RFID tags to authenticate and confirm the visitor's identity card.

A concept of global addressing and integration of 6LowPAN with sensor devices such as ZigBee sensors (802.15.4) and other technologies which lack IPv6 capabilities in their stack, has introduced in [16] as Glowbal IP. The main part of Glowbal IP is Access Address Identifier (AAID), which plays the role of header. AAID contains the IPv6 and UDP parameters (e.g. source/destination ports, source/destination address) to reduce the overhead in IPv6 and 6LowPAN. AAID gateway contains a Local to Global mapping table (L2G) in order to save the mapping process. Glowbal IPv6 provides the integration between most technologies (e.g. Konnex, enabling IPv6 for smart phones through their Bluetooth Low Energy interface, ZigBee and so on) and optimizes 6LowPAN by reducing 41 bytes of IPv6/UDP headers. Glowbal IPv6 protocol uses the DNS-SD (DNS service Directory) and (m-DNS) multicast DNS for the discovery services, which are developed from DNS to enable Glowbal IPv6 for legacy technologies such as X10 and Konnex. This protocol is not suitable for devices such as smart phones, as this feature is not implemented for them.

An IPv6 addressing mechanism to address legacy technologies has been introduced in [17]. The aim of this mechanism is allocating IPv6 to technologies that do not support this protocol in order to increase the number of connected devices and technologies to IoT. In addition, there will be interoperability among different devices (e.g. sensors). Legacy technologies that benefit from this mechanism include European Installation Bus (EIB) [98], Controller Area Network (CAN) for building automation, and RFID for identification. In this method, IPv6 addressing proxy is provided for mapping devices in a way that the device ID is used to create host ID, which will be combined with net ID for obtaining the IPv6. The IPv6 mapping mechanism is applied to devices and technologies using their data frame to obtain a mapped frame.

Authors in [18] have focused on the connection between the Internet of Things and cloud, using IPv6 and Constrained Application Protocol (CoAP). The presented platform benefits from Software as a Service (SaaS), Universal Device Gateway (UDG), and IoT6 project to allocate IPv6 addresses to various devices. The platform presented in [18] solved the cloud security issues through UDG project. In addition, it used ZigBee and 6LowPAN for IPv6 adaption in IoT devices.

Naming, Addressing and Profile Server (NAPS) middleware presented in [19] makes different platforms interoperable in IoT sensory networks. This method covers several protocols for addressing such as RFID, ZigBee, Bluetooth, etc. The IPv6 address conversion parameter provided in [19] is as follows:

$$\underbrace{Address-1@protocol-1/address-n@protocol-n}_{\text{none IP networks}}/\text{IP-address}$$

As can be seen, protocols that do not have the advantage of IPv6 are converted by the first section and if an IPv6 already exists, it will be used directly. An example of this mechanism for RFID using EPC is [19]:

binary-epc@rfid/<readerIP:port>

A Tree-Code model for addressing the non-ID physical objects has been presented in [20]. The purpose of this mechanism is connecting non-ID physical objects to the Internet. The focus of this mechanism is on non-ID physical objects, but it also works on objects that have an ID (e.g. EPC). The proposed mechanism in [20] operates based on the properties of things such as location (if GPS exists), physical characteristics (e.g. color, size, shape and so on), and their behavior (their interaction with environment and other things).

A simple and useful EPC based mechanism has been reported in [21]. The mechanism uses serial part of EPC (which is the most unique part and it is always less than or equal to 64 bits) for the mapping process. This method adds one padding, if required, for completing the 64 bits. Then, the obtained 64 bits will be converted to hexadecimal. Therefore, IPv6 address construction is done by combining it with the reader net ID.

### III. THE RFID MEMORY STRUCTURE

RFID system has a component called the memory, which consists of memory bank (MB), where each of these MBs accommodate information about an application or purpose. Fig. 3 clearly demonstrates all the details. The structure that can be observed in the figure is common across EPC Gen2 and ISO/IEC 18000-6 and ISO/IEC 18000-3 Mode 3 [22]. TID and UII each located in MB $10_2$ and $01_2$, respectively. MB$10_2$ and $01_2$ are the parts which we need, since TID and UII located in them. The values of MB $10_2$ are determined by the 8 bits Allocation Class (AC) identifier provided by ISO / IEC 15963[23, 24]. If the value of this field is $11100010_s = E2_h$, then it indicates the presence of the TID with the EPC standard structure. The following fields which exist based on this standard are 9 bits Tag Mask Designer Identifier (Tag MDID), and 12 bits tag model number. But in some cases, the extended tag identification number (XTID)/serial number field which is 48 bits or more, is also visible. ISO is another standard with $11100000_2=E0_h$ of AC identifier value which takes place of EPC as TID. Next fields that are based on this standard include 8 bits manufacturer identifier and 48 bits (or more) tag serial number [25, 26]. Fig. 2 presents the TID memory structure for each of the mentioned standards. On the other hand, UII forms when the tag is incorporated in the object, thus it is called Birth Record and MB$01_2$ is called Name, which is unique [27, 28, 29]. UII format is identified by $17_h$ bit, which is called Toggle bit. If this bit is 0, then it suggests presence of EPC as UII tag. On the other hand, if this bit is 1, then UII follows the standard ISO format. In this state, 18h bit known as Applicability Family Identifier (AFI) has values with a length of 1 byte, which considers different schemes for the ISO UII part. AFI has special values to develop difference between different applications. Any type of AFI values has been defined by ISO organization for specific purposes, and values defined for supply chain standards are A1, A2, A3, A4, A5, etc. [30]. The importance of presence of toggle is further highlighted in the next section. Fig. 4 has presented the structure of MB$01_2$.

For some standards, UII has a Data Identifier (DI). DI is a predetermined character (or an array of characters) which defines general classification or use of the information of interest. A type of DI which has been defined for unique identification of items is 25S. Usage of DI25S suggests that the data string has been uniquely allocated to identification of one item. The data string 25S consists of 18V part and the supplier serial number. The serial number related to the supplier is totally unique and created by 18V. However, 18V part itself is composed of issuing agency code (IAC) related to ISO/IEC 15459 and Company Identification Number (CIN). The serial number part includes a unique serial number for the CIN in 18V. The serializing methods of part number, and/or lot/batch in the serial number part consist of the two following methods:

part number + serial number (unique for that part number for the CIN)

lot/batch number + serial number (unique within the lot/batch for the CIN)

Another DI which can be mentioned is 25B. Indeed, 25B has been defined in ISO 17364, which includes the CIN and Issuing Agency Code (IAC) in accordance with ISO/IEC 15459 sections. The DI 25B structure is as follows:

IAC, followed by CIN and the RTI serial number which is unique within the CIN holder's domain [31]. Typically, UII accommodates values with 25S format [32]. Fig. 1 provides an example of UII with its URN form.

| DI | IAC | Company Identification Number (CIN) | Serial Number |
|---|---|---|---|
| | | | |
| urn:iso:std:iso-iec:15459:25S.UN.15849561.6920001026 | | | |

| DI | IAC | Company Identification Number (CIN) | Part Number | Serial Number |
|---|---|---|---|---|
| | | | | |
| urn:iso:std:iso-iec:15459:25S.UN.11639768.MH26231276 | | | | |

Figure. 1. Examples of ISO/IEC 15459 UII format and URN structure [22]

| TID MB Bit Address | BIT Address (In Hex) | | | | | | | | | | | | | | | |
|---|---|---|---|---|---|---|---|---|---|---|---|---|---|---|---|---|
| | 0 | 1 | 2 | 3 | 4 | 5 | 6 | 7 | 8 | 9 | A | B | C | D | E | F |
| $10_h$-$1F_h$ | Tag MDID (last 4-bits) | | | | Tag Model Number (12-bits) | | | | | | | | | | | |
| $00_h$-$0F_h$ | $11100010_2$=$E2_h$ | | | | | | | | Tag MDID (first 8 bits) | | | | | | | |

(a)

| MB Bit Address | 0 | 1 | 2 | 3 | 4 | 5 | 6 | 7 | 8 | 9 | A | B | C | D | E | F |
|---|---|---|---|---|---|---|---|---|---|---|---|---|---|---|---|---|
| $30_h$-$3F_h$ | Tag Serial Number (48 bits) | | | | | | | | | | | | | | | |
| $20_h$-$2F_h$ | | | | | | | | | | | | | | | | |
| $10_h$-$1F$ | | | | | | | | | | | | | | | | |
| $00_h$-$0F_h$ | $11100000_2$=$E0_h$ | | | | | | | | Tag Manufacturer ID(first 8 bits) | | | | | | | |

(b)

Figure. 2. RFID $MB10_2$ structure [26] (a) TID memory structure in the current EPC standards. (b) TID memory structure in the ISO standards.

| TID | BIT Address (In Hex) |
|---|---|

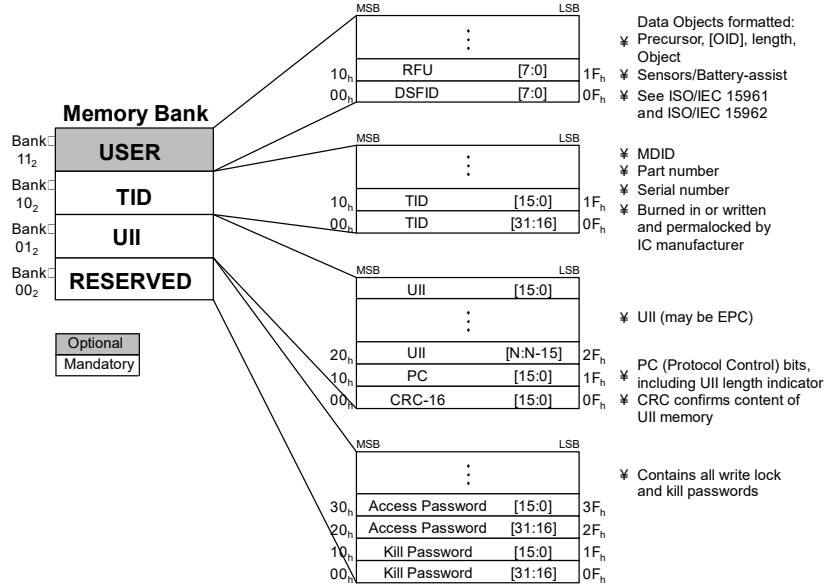

Figure. 3. RFID Memory Banks for ISO 18000-6 and EPC Gen.2 [22]

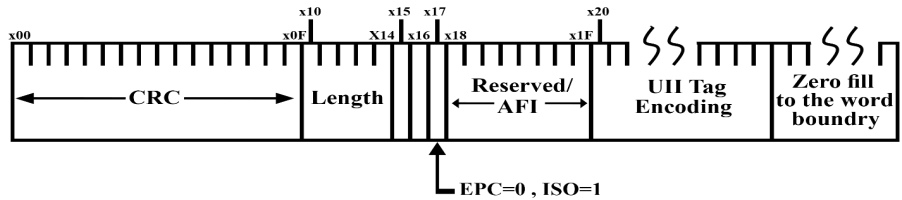

Figure. 4. RFID $MB01_2$ structure [22]

## IV. PROPOSED MECHANISM

As mentioned previously, giving address to the objects to be connected to the Internet requires special processes. Meanwhile, RFID systems have different standards, and the addressing mechanism related to it should be common across all different types of RFIDs, such that it develops interoperability among tags across different environments. Further, the addressing of tags should be optimal, fast, scalable, inexpensive, and simple. A new method is presented further for this purpose.

The presented mechanism benefits from $MB01_2$ which has values such as toggle and UII. The proposed mechanism benefits from an algorithm which improves the addressing, as introduced further.

### The selection algorithm

When EPC schemes are changed into binary, the number of their bits may exceed 64. Since we only need 64 bits for merging with Net ID section to create an IPv6 address, through a technique, only 64 bits of the entire obtained binary domain should be used. Various mechanisms use compression methods and hash functions. The authors in [33] have thoroughly examined this type of methods and concluded that usage of compression and hash functions requires complex computations and a great deal of time. Therefore, in this paper we propose a selection algorithm which does not require complex computations and is not time-consuming either. It is assumed that our binary numbers located in an initial list. Therefore, regarding the operation of the algorithm, first it investigates the number of bits, and if the number of bits exceeded 64, it selects the last 64 bits which are valuable and unique bits in EPC schemes from the initial list and embeds it in a new list called selection list. Eventually, it outputs the list for the main mechanism and uses it. Evidently, this algorithm is very simple and has a low time complexity, and needs short processing time. The rest of the addressing mechanism is explained further.

Regarding the addressing mechanism, first the toggle bit is read and based on its contents, the type of RFID tag UII is identified. If the toggle was equal to 0, then the standard is EPC, where its entire structure is converted to binary, which results in three possible states: less than 64 bits, equal to 64 bits, and larger than 64 bits. If the first state happened, zero is added prior to the binary numbers. If the second state happened, then the same 64 bits are used without any further manipulation. However, if the third state happened, the algorithm chooses 64 bits from the entire binary domain to continue the addressing operations, as explained above. Nevertheless, if the toggle was 1, it represents a UII which supports ISO format. The format of each ISO UII is specified by AFI bit, which varies given its type of application. However, in some of its syntaxes, it has character values, whose reading requires further processing and hence is time-consuming. On the other hand, any syntax has constantly a serial number value. Therefore, the serial number section is read and its value is converted to binary (the value of this section is either decimal or hexadecimal). Then, as with EPC section, its length is judged. Unlike the EPC state, it always follows the states of either less than 64 bits or equal to 64 bits, since the length of a serial number is at most 64 bits. In the first state, one padding operation is performed, while in the second state the serial number is used without further manipulation. Then, 64 bits of Net ID is added to the serial number, whereby eventually the main format of the IPv6 address is created which is unique and hierarchical.

The second addressing method focuses on $MB10_2$ namely TID memory which is already discussed in details in pervious sections. The mechanism initials by reading the AC field and decides based on its value. If the content of this field is equal to $E2_h$, then the three Tag MDID, Tag model number and XTID (if exists) fields are used for addressing. When these fields are converted to binary, three states of less than 64 bits, equal to 64 bits and greater than 64 bits will happen. In the first state, left zero padding is implemented. If the second state happened, the same 64 bits is used without any manipulation and in some cases if the third state happened, 64 bits are selected and used. On the other hand, if the value of AC field equals to $E0_h$ thus the standard is ISO. Therefore, only a serial number field is read and its value is used. which has three states same as the previous standard. With the difference that the occurrence probability of the third state is low, thus the selection algorithm is not used either.

Consequently, the result of each operation above is combined with the 64 bits of reader Net ID and generates a new IPv6 which is unique and hierarchical too. Fig. 5 presents the python implementation of proposed mechanisms. At first it detects the toggle bit then, a unique IPv6 address will be generate based on ISO/EPC standard. Also the executed time for generating a unique IPv6 is about 10 milliseconds which is a very short time for address generation in comparison with other similar methods.

```
The Toggle Bit is:  0
RFID Standard is EPC:
961186085415459865490825641692369
your Unique IPv6 is :
5490:8256:4169:2369:6789:1011:1213:1415 /128
Generation Time is:  0.007996082305908203 /S

The Toggle Bit is:  1
The RFID Standard is ISO:  9611860854
Your Unique IPv6 is :
0000:0096:1186:0854:6789:1011:1213:1415 /128
Generation Time is:  0.006995677947998047 /S
```

Figure. 5. Results of proposed addressing methods.

## V. CONCLUSION

In this paper, attempts were made to present unique methods for giving address to objects connected to the Internet of Things. The aim of this paper was to cover different RFID standards for the mapping operations. This paper presented general mechanisms including two efficient algorithms in order to facilitate and accelerate the operations of mapping their addresses. These are the first methods presented for supporting the supply chain completely. The most important characteristics of the proposed methods include simplicity, hierarchy, high interoperability, scalability, low implementation cost, and performing the computations within a short time.